
\overfullrule=0pt
\magnification=1200 
\baselineskip=5ex
\raggedbottom
\font\fivepoint=cmr5
\headline={\hfill{\fivepoint  EHLMLRJM-22/Feb/93}}

\def\uprho{\raise1pt\hbox{$\rho$}}
\def\mfr#1/#2{\hbox{${{#1} \over {#2}}$}}
\catcode`@=11
\def\eqalignii#1{\,\vcenter{\openup1\jot \m@th
\ialign{\strut\hfil$\displaystyle{##}$&
        $\displaystyle{{}##}$\hfil&
        $\displaystyle{{}##}$\hfil\crcr#1\crcr}}\,}
\catcode`@=12
\def\Tr{{\rm Tr}}
\def\1{{\bf 1}}
\def\endproof{\hfill QED. \bigskip}
\def\Trace#1{\Tr \left[ #1 \ebH \right]}
\def\acomm#1,#2{\left\{#1,#2\right\}}
\def\rhobe{\uprho_\beta}
\def\rhobes{\uprho_{\beta\sigma}}
\def\rhoN{\rhobe^{(n)}}
\def\rhosN{\rhobes^{(n)}}
\def\ebH{e^{-\beta H}}

\def\Boltzf#1{(1+e^{\beta\lambda_{#1}})^{-1}}
\def\abs#1{\vert #1 \vert}

\def\cd#1{c^\dagger_{#1}}
\def\c#1{c^{\phantom\dagger}_{#1}}
\def\bd#1{b^\dagger_{#1}}
\def\b#1{b^{\phantom\dagger}_{#1}}
\def\bdl#1{\bd{\lambda_{#1}}}
\def\bl#1{\b{\lambda_{#1}}}

\def\bm#1{\b{\mu_{#1}}}
\def\nl#1{n_{\lambda_{#1}}}
\def\contr#1,#2{\kern+1ex\overline{\phantom{\bd{}b}}\kern-3.5ex #1#2}

\def\detij#1{{\rm det} \left[ #1 \right]_{i,j=1}^n}
\def\deltadet{\detij {\delta_{x_i y_j}}}
\def\deltadett{\detij { \delta_{x_i y_j} \delta_{\sigma_i \tau_j} } }
\def\spinstates{\{\uparrow,\downarrow\}}
\centerline{\bf UNIFORM DENSITY THEOREM FOR THE HUBBARD MODEL}
\smallskip
\centerline{(to appear in Journal of Mathematical Physics, March 1993)}
\bigskip
\bigskip
\bigskip\noindent
Elliott H. Lieb\footnote*{E-mail: lieb@math.princeton.edu}

\noindent
{\it Departments of Physics and Mathematics, Princeton University,
P.O. Box 708, Princeton, NJ 08544-0708}

\bigskip\noindent
Michael Loss\footnote\dag{E-mail: loss@math.gatech.edu}

\noindent
{\it School of Mathematics, Georgia Institute of Technology,
Atlanta, GA  30332-0160}

\bigskip\noindent
Robert J. McCann\footnote\ddag{E-mail: mccann@math.princeton.edu}

\noindent
{\it Department of Mathematics, Princeton University, Princeton,
NJ 08544-1000}
\bigskip
\bigskip\noindent
{\bf Abstract:}
A general class of hopping models on a finite bipartite lattice is
considered, including the Hubbard model and the Falicov-Kimball model.
For the half-filled band, the single-particle density matrix $\uprho
(x,y)$ in the ground state and in the canonical and grand canonical
ensembles is shown to be constant on the diagonal $x=y$, and to vanish
if $x \not=y$ and if $x$ and $y$ are on the same sublattice. For free
electron hopping models, it is shown in addition that there are no
correlations between sites of the same sublattice in any higher order
density  matrix. Physical implications are discussed.

\bigskip\bigskip\bigskip

The one-particle reduced density matrix, $\uprho (x,y)$, of a
many-electron quantum system can reveal a good deal about the presence
or absence of spatial uniformity.  For a general class of tight
binding models,  including the Hubbard model and the Falicov-Kimball
model,  $\uprho (x,y)$ turns out to have a particularly simple form
in the case of the half-filled band.  The result, which is applicable
to a bipartite lattice, is that $\uprho (x,x)$ is always exactly equal
to one for all $x$ on a finite lattice, even though {\it the hopping matrix
elements and interaction are nonuniform, random and uncorrelated}.
Furthermore $\uprho (x,y) = 0$ when $x \not= y$ but $x$ and $y$ are in
the same sublattice.  This fact has been long appreciated in the
chemistry literature,  in the context of certain models describing
$\pi$ electrons in conjugated carbon systems$^{1}$.  It was
first observed for the H\"uckel (free electron) model$^{2}$ by
Coulson and Rushbrooke$^3$,  who
used it to justify the assumption that the effective potential should
be the same at each carbon site in a self-consistent molecular orbital
treatment.  MacLachlan$^{4,5}$ extended the result via a hole-particle
symmetry argument to the Pariser-Parr-Pople (interacting electron)
model$^{6,7}$,  of which
the Hubbard model is formally a special case.  For so-called
{\it alternant} molecules,  i.e., those in which the carbon atoms form
a bipartite lattice,  MacLachlan showed that to each $N$ electron
eigenstate corresponds a $2\abs\Lambda-N$ electron eigenstate with the
complementary density.  (Here $\abs\Lambda$ is the number of carbon sites,
and there are 2 allowed spins at each site).
These states have the same energy, up to a shift
which is proportional to $\abs\Lambda-N$.  He used this
{\it pairing theorem} to explain the identical spectra which had been
observed for positive and negative ions of the same alternant
molecule.  Despite its usefulness,  the result has remained
unknown in the statistical mechanics and solid state physics
literature,  where the same models are used to investigate phase
transitions and the existence of long range order.  In this context,
the persistence of uniformity in the face of randomness is striking.
For instance, it hints at the stability of
some periodic structures in solids, which is to say that the
occurrence of periodic structures in a system might be insensitive to
some of the details of the Hamiltonian of the system.  We therefore
present the theorem in the context of statistical mechanical ensembles
at positive and zero temperature,  together with an extremely simple proof.
Along the way,  we extend the applicability of the
theorem to include Hamiltonians with spin dependent hopping and
spin-flip interactions.  Thus consequences are extracted for some
additional models of current physical interest, such as the
Falicov-Kimball model.  Finally,  the infinite volume limit is discussed;
if the infinite volume Gibbs state has non-constant density then it is
not unique, and the theorem guarantees the existence of another Gibbs state
having the complementary density.

To establish some notation, we consider a finite graph (lattice)
consisting of sites labeled by $x, y$, etc. and edges (or bonds)
connecting certain pairs of sites.  We assume that the graph is
{\it bipartite}, i.e., the vertices can be divided into two disjoint subsets
$A$ and $B$ such that there is no edge connecting $x$ and $y$ if $x$
and $y$ are both in $A$ or both in $B$.  The total number of sites in
$\Lambda, A$ or $B$ is denoted by $\vert \Lambda \vert, \vert A \vert$
or $\vert B \vert$.  We assume that $\vert A \vert \geq \vert B
\vert$.  We are given a hermitian $\vert \Lambda \vert \times \vert
\Lambda \vert$ hopping matrix \ $T$ with elements $t_{xy}^{\phantom{*}}
= t_{yx}^*$.  These elements are nonzero only if $x$ and $y$ are
connected by an edge; thus the elements $t_{xy}$ are zero
whenever $x$ and $y$ are both in $A$ or both in $B$. (In particular,
$t_{xx}=0$ for all $x$.) Physically, $T$ originates in overlap integrals
and is real in the absence of magnetic fields that interact with the
electron orbital motion.

It is easy to see that the nonzero eigenvalues of $T$ come in opposite
pairs: for every eigenvalue $\lambda$ there is an eigenvalue
$-\lambda$.  The two corresponding eigenvectors $\phi^\lambda$ and
$\phi^{-\lambda}$ are conjugates in the following sense:  if
$\phi^\lambda = (f^\lambda, g^\lambda)$ with $f^\lambda$ being the
$A$-part of $\phi^\lambda$ and $g^\lambda$ being the $B$-part of
$\phi^\lambda$ then $\phi^{-\lambda} = (f^\lambda, - g^\lambda)$.
Alternatively, if $V = \pmatrix{1&0\cr0&-1\cr} = V^\dagger$ is the
diagonal unitary matrix that multiplies $\phi$ by $-1$ on the $B$-
sites, then $VTV = -T$.  It is possible that $T$ has zero-modes;
indeed if $\vert A \vert > \vert B \vert$ then $T$ will have at least
$\vert A \vert - \vert B \vert$ zero-modes and all these $\lambda = 0$
eigenvectors have the form $\phi = (f,0)$.

Suppose now that we have a half-filled band, i.e., $N = \vert \Lambda
\vert$ electrons.  By virtue of the two spin states for each electron
we have that the ground state energy of $H_0 = \sum\nolimits_{x,y}
t_{xy} c^\dagger_x c_y$ is $E = 2 \sum\nolimits_{\lambda < 0}
\lambda$.  (Note:  If $\vert \Lambda \vert$ is odd there is at least
one zero-mode, and so this formula is correct even in this case.) \ The
ground state of $H_0$ might be degenerate, however (because of
zero-modes).  We {\it define} the density matrix for spin $\sigma$ in
the ground state to be
$$\uprho_\sigma (x,y) = \sum \limits_{\lambda <
0} \phi^\lambda (x)^* \phi^\lambda (y) + \mfr1/2 \sum \limits_{\lambda
= 0} \phi^\lambda (x)^* \phi^\lambda (y).  \eqno(1)$$
We see that $\Tr
\ \uprho_\sigma = \sum \nolimits_x \uprho_\sigma (x,x) = \vert \Lambda
\vert /2$, as it should, and that $\uprho_\sigma$ agrees with the
$\beta \rightarrow \infty$ limit of the positive temperature density
matrix, defined in the {\it grand canonical} ensemble by
$$\uprho_{\beta \sigma} (x,y) = \sum \limits_\lambda \phi^\lambda (x)^*
\phi^\lambda (y) \, e^{-\beta \lambda}/(1 + e^{-\beta \lambda}).
\eqno(2)$$
Note that we have used zero chemical potential which, by
virtue of the $\lambda, - \lambda$ symmetry, always yields $\vert
\Lambda \vert$ as the average particle number.  If there are
zero-modes the ground state, and the $\uprho_\sigma$ in the ground
state, will not be unique. Eq. (1) serves to fix $\uprho_\sigma$ for our
purposes.  The Gibbs state is always unique for a finite volume.

Another Gibbs state with which we shall be concerned is the {\it
canonical} ensemble.  The density matrix here will be denoted by
$\widetilde {\uprho}_{\beta \sigma} (x,y)$.  Its definition is well
known and we shall not write it explicitly for $H_0$, but we note that
the $\beta \rightarrow \infty$ limit of $\widetilde{\uprho}_{\beta
\sigma}$ also equals $\uprho_\sigma$.

Coulson and Rushbrooke's$^{3}$ observation regarding (1) is the starting
point of our further analysis.  If $x\in A$ and $y \in A$ then, using
the fact that $\phi^\lambda (x) = \phi^{-\lambda} (x)$, we have that
$$\uprho_\sigma(x,y) = \mfr1/2 \sum \limits_{{\rm all} \ \lambda}
\phi^\lambda (x)^* \phi^\lambda (y) = \mfr1/2 \delta_{xy}$$
since the $\phi^\lambda$'s
form an orthonormal basis.  A similar remark holds for $x,y \in B$.
Thus, $\uprho_\sigma (x,x) = \mfr1/2$ for all $x \in \Lambda$ and
$\uprho_\sigma (x,y) = 0$ for $x,y \in A$ or $x,y \in B$.  As we shall
see from
the following general theorem (by specializing to zero interaction) the
same conclusion applies to $\uprho_{\beta \sigma} (x,y)$ and
$\widetilde{\uprho}_{\beta \sigma} (x,y)$.

\bigskip
\centerline{\bf THE GENERALIZED HUBBARD MODEL}

The interacting system we shall be concerned with is the {\it
generalized} Hubbard model defined by the Hamiltonian (with spin
dependent hopping) $$H = \sum\limits_\sigma \sum \limits_{x,y \in
\Lambda} t^{\phantom{\dagger}}_{xy\sigma} c^\dagger_{x\sigma}
c^{\phantom{\dagger}}_{y \sigma} + \sum \limits_{\sigma,\tau} \sum
\limits_{x,y \in \Lambda} U_{xy\sigma\tau} (2n_{x \sigma} -1)
(2n_{y\tau} -1), \eqno(3)$$ where $n_{x \sigma} = c^\dagger_{x\sigma}
c^{\phantom{\dagger}}_{x \sigma}$ and with $U_{xy\sigma\tau}$ real (but
not necessarily of one sign).  $T_\sigma = \{ t_{xy\sigma}\}$ is
hermitian and bipartite for each $\sigma = \uparrow$ or $\downarrow$.
If we take $T_\uparrow = T_\downarrow$ and $U_{xy\sigma \tau} = U
\delta_{xy}$ then $H$ is the usual Hubbard Hamiltonian (apart from a
trivial {\it additive} constant) with interaction $8U \sum
n_{x\uparrow} n_{x \downarrow}$.  The noninteracting case, $H_0$,
corresponds to $U_{xy \sigma\tau} = 0$.  In general, the total spin
angular momentum ($SU(2)$ symmetry) will be conserved if we require
$t_{xy\sigma}$ and $U_{xy\sigma\tau}$ to be independent of the spin
labels $\sigma$ and $\tau$; for our purposes we do not require this
$SU(2)$ invariance.

The positive temperature, grand canonical density matrix $\uprho_{\beta
\sigma}$ is defined to be $$\uprho_{\beta \sigma} (x,y) = Z^{-1}\Tr
[c^\dagger_{x \sigma} c^{\phantom{\dagger}}_{y\sigma} e^{-\beta
H}], \eqno(4)$$
where $Z=\Trace{}$.  Formula (4) reduces to (2) for the
noninteracting case.  The trace is over the full Fock space containing
all particle numbers ranging from 0 to $2 \vert \Lambda \vert$.  Again,
the zero chemical potential in (4) insures that $\Tr \ \uprho_{\beta
\sigma} = \vert \Lambda \vert /2$.

The canonical density matrix $\widetilde{\uprho}_{\beta \sigma} (x,y)$
for this model is also given by (4), but where the trace is only over
the $N$-particle sector (note that both $H$ and $c^\dagger_{x \sigma}
c^{\phantom{\dagger}}_{y \sigma}$ leave this sector invariant).  The
half-filled band is defined by $N = \vert \Lambda \vert$.  Since $(\Tr
ABC)^* = \Tr \ C^\dagger B^\dagger A^\dagger = \Tr \ B^\dagger
A^\dagger C^\dagger$ we see that $\uprho_{\beta \sigma}$ and
$\widetilde{\uprho}_{\beta \sigma}$ are hermitian matrices for each
$\sigma$.

\proclaim
THEOREM: (Uniform Density in the Generalized Hubbard Model).
\par
\noindent
{\it The canonical and the grand canonical density
matrices satisfy: }
$$\eqalignno{\widetilde{\uprho}_{\beta\sigma} (x,x)
= \uprho_{\beta \sigma} (x,x) = 1/2 \quad &\hbox{for all} \ x \in
\Lambda \qquad&(5)\cr \widetilde{\uprho}_{\beta \sigma} (x,y) =
\uprho_{\beta \sigma} (x,y) = 0\phantom{/2} \quad &\hbox{if} \ x,y \in
A \ \hbox{or} \ x,y \in B. \qquad&(6)\cr}$$
\par
\noindent
{\it Proof:}  The proof for $\widetilde{\uprho}_{\beta\sigma}$ will be
the same as that for $\uprho_{\beta \sigma}$ so we shall only give the
proof for the latter.

First, we consider real $T$.  MacLachlan's version$^{5}$ of the
hole-particle unitary transformation, $$c^{\phantom{\dagger}}_{x\sigma}
\leftrightarrow c^\dagger_{x \sigma} \ {\rm for} \ x \in A, \qquad
c^{\phantom{\dagger}}_{x \sigma} \leftrightarrow -c^\dagger_{x \sigma}
\ {\rm for} \ x \in B, \eqno(7)$$ evidently leaves the Hamiltonian $H$
and the relevant Hilbert spaces invariant.  If this unitary
transformation is denoted by $W$ we have that $W^2 = 1$, $W =
W^\dagger$, $WHW = H$ and hence $Z
\uprho_{\beta\sigma} (x,y) = \Tr [W c^\dagger_{x\sigma}
c^{\phantom{\dagger}}_{y\sigma} WW e^{-\beta H} W] = \Tr [(W
c^\dagger_{x\sigma} W) (Wc_{y\sigma}W) e^{-\beta H}]$.  If $x,y \in A$
we can use (7) and the fermion commutation rule to conclude that
$\uprho_{\beta \sigma} (x,y) = \delta_{xy} - \uprho_{\beta \sigma}
(y,x)$.  The same is true if $x,y \in B$.  If $T$ is real,
$\uprho_{\beta\sigma}$ is evidently real; since $\uprho_{\beta \sigma}$
is also hermitian the theorem is proved in the real case.

The complex case is a bit subtle.  The Hamiltonian $H$ is no longer
invariant under the hole-particle transformation $W$, but it is
invariant under the {\it antiunitary} transformation $Y = JW$, in which
$J$ is complex conjugation.  More precisely, any vector $\Psi$ in our
Hilbert space can be written as a linear combination, with complex
coefficients, of the basis vectors consisting of monomials in the
$c^\dagger_{x\sigma}$'s applied to the vacuum.  The antiunitary map $J$
acts on $\Psi$ by replacing each coefficient by its complex conjugate.
We note that $JW = WJ$ and therefore $Y^2 = \1$.  It is also easy to
see that $YHY = H$ and that $Y c_{x\sigma} Y = W c_{x\sigma} W$, which
is given by (7).

Now suppose that $K$ is an arbitrary linear operator, and consider $L
\equiv JKJ$.  Although $J$ is nonlinear, it is easy to check that $L$
is linear.  In fact the matrix elements of $L$ in the above mentioned
basis are simply the complex conjugates of the corresponding elements
of $K$.  Therefore, even though $J^2 = 1$, it is {\it not} generally
true that $\Tr L = \Tr JKJ = \Tr KJ^2 = \Tr K$.  What is true is that
$\Tr L = (\Tr K)^*$.

In our case we have, for $x,y \in A$ or $x,y \in B$, \ $Z
 \uprho_{\beta\sigma} (x,y)^* = \Tr [JW
c^\dagger_{x\sigma} c^{\phantom{\dagger}}_{y\sigma} e^{-\beta H} WJ]
\hfill\break = \Tr [(Y c^\dagger_{x\sigma} Y)
(Yc^{\phantom{\dagger}}_{y\sigma}Y) (Ye^{-\beta H} Y)] = \Tr
[c^{\phantom{\dagger}}_{x\sigma} c^\dagger_{y\sigma} e^{-\beta H}] =
Z \left( \delta_{xy} - \uprho_{\beta\sigma} (y,x) \right)$.  The
hermiticity of $\uprho_{\beta\sigma}$ now implies the theorem.
\endproof

It is worth noting that only the invariance of $H$ under $Y$ and the
bipartite structure of the lattice have been used.  Thus the theorem
(with some obvious modifications) applies to the case where the kinetic
energy has spin-flip terms and is given by
$$\sum\limits_{x,y,\sigma,\tau} t^{\phantom{\dagger}}_{xy\sigma\tau}
c^\dagger_{x\sigma} c^{\phantom{\dagger}}_{y\tau}. \eqno(8)$$
Hermiticity requires $t_{xy\sigma\tau} = t^*_{yx\tau\sigma}$;
to be a hopping model on a bipartite lattice imposes another condition on
$T$ which will be made clear in a moment. To apply the previous theorem,
we observe that this model is equivalent
to a system of spinless fermions living on a lattice twice as large as
$\Lambda$, with new co-ordinates $(x, \sigma)$,  where $x \in \Lambda$ and
$\sigma \in \spinstates$.  The new lattice should be bipartite,  which
is to say that it can be divided into two sublattices such that
$t_{xy\sigma\tau}=0$ when both $(x, \sigma)$ and $(y,\tau)$ are in the
same sublattice. $\uprho_{\beta\sigma} (x,y)$ is replaced by
$\rhobe (x, \sigma; y, \tau) = \rhobe (y, \tau; x, \sigma)^*$, and the
extended theorem states that it equals
$\mfr1/2\delta_{xy} \delta_{\sigma\tau}$ when $(x,\sigma)$ and $(y,\tau)$
are both on the same sublattice.  In practice,  the only subtlety is the
identification of the sublattices.  For example,  let $\Lambda$
be decomposed into $A$ and $B$.  Then the theorem applies if
$t_{xy\sigma\tau}=0$ whenever
$x,y \in A$ or $x,y \in B$.  In this case one of the sublattices would
consist of those sites $(a, \sigma)$ for which $a \in A$ and
$\sigma \in \spinstates$.   Alternatively,  we could have taken the
condition to be $t_{xy\sigma\sigma}=0$  when $x,y \in A$ or $x,y \in B$,
and $t_{xy\uparrow\downarrow}=0$ whenever one of $x,y$ is in $A$ and the
other in $B$.  This model also satisfies the hypothesis,  but in this case
one of the sublattices
consists of $\left\{ \;(a,\uparrow) \;|\; a \in A \;\right\}
  \,\bigcup\,
 \left\{ \; (b,\downarrow) \; | \; b \in B \;\right\}$.
The latter scenario would be realized if we had only on-site spin-flip
terms: $t_{xy\uparrow\downarrow}=t_x\delta_{xy}$, a physically appealing
possibility.  In the interest of simplicity of notation and exposition
we have relegated this generalization --- relevant to some physical
models --- to the remark here.

\def\Sz{S^{3}_{x}}
\def\Splus{S^+_x}
\def\Sminus{S^-_x}
\def\S{\vec {S_x}}
The theorem also extends to include spin-spin interactions of a more
general nature than those manifested in (3).  This is particularly welcome,
since many physical models --- the t-J model for example --- require
SU(2) invariant spin couplings.  Let $x \in \Lambda$ and
consider the operators
$\Sz = (n_{x\uparrow} - n_{x\downarrow})/2$
and $\Splus = \cd{x\uparrow}\c{x\downarrow}$.  They generate the usual
SU(2) algebra at site $x$:
$$\S = \left(\,(\Splus + \Sminus)/2, \,(\Splus - \Sminus)/2i, \,\Sz \right)
$$
where $\Sminus = (\Splus)^\dagger$ and the total spin at $x$ is
$\S\cdot\S$.  In terms of these operators, the spin-spin
interaction realized in (3) is of the form $\Sz S^3_y$ where
$x, y \in \Lambda$;  it occurs
with a coupling constant proportional to
$(U_{xy\uparrow\uparrow} + U_{xy\downarrow\downarrow}) -
              (U_{xy\uparrow\downarrow} + U_{xy\downarrow\uparrow})$.
Since $\Sz \leftrightarrow -\Sz$ under the hole-particle inversion $Y$,
the invariance of this interaction is
manifest,  but it clearly destroys the SU(2) symmetry of the Hamiltonian.
However,  the antiunitary transformation $Y$ does not single out the
$3$-direction;  indeed it follows
that $\S \leftrightarrow -\S$ under $Y$.  As a result,    the
conclusions of the uniform density theorem,  (5) and (6),  are
valid for any system described by a Hamiltonian
$$H' = \sum\limits_\sigma \sum \limits_{x,y \in
\Lambda} t^{\phantom{\dagger}}_{xy\sigma} c^\dagger_{x\sigma}
c^{\phantom{\dagger}}_{y \sigma} +
\sum\limits_{x,y \in \Lambda} U_{xy} (n_x - 1)(n_y - 1)  +
\sum\limits_{x \ne y} \sum\limits_{i,j=1}^3
J_{xy}^{ij} S^i_x S^j_y. $$
Here $n_x = n_{x\uparrow} + n_{y\downarrow}$ and
$(S_x^1, S_x^2, S_x^3) = \S$,
$t_{xy\sigma}$ is as in (3) and the $U_{xy}$ and $J_{xy}^{ij}$
($i,j=1,2,3$) are completely arbitrary real constants.  The case
$J_{xy}^{ij}=J_{xy}\delta_{ij}$ corresponds to the
SU(2) invariant interaction $\S \cdot \vec{S_y}$.  We have
omitted the spin-charge couplings $(n_x-1)S^i_y$, which would appear
in the most general form of $H'$, in order to maintain its physical
simplicity.

We conclude by discussing some consequences and extensions of the
Theorem in certain limiting cases.

{\bf I.  No Correlations within a Sublattice for the Free Electron Model:}

For the free electron model,  the interaction terms $U_{xy\sigma\tau}$
are zero, leaving a
purely kinetic Hamiltonian of the form (3) --- or (8),  although we
continue to suppress this generalization for notational convenience.
This Hamiltonian is
quadratic in the creation and annihilation operators,  i.e.,  it is a
one-body operator.  Wick's theorem$^{8}$ therefore applies to
the grand canonical ensemble$^{9}$, for which it states that all
operator product expectations are expressible in terms of pair
expectations.  We use this observation together with the vanishing of
$\rhobes (x, y)$ for $x,y$ in the same sublattice to show that there
can be no correlations
between the electron densities on sites in this sublattice. This
striking result does not seem to appear
previously in the literature,  although it generalizes a comment
made on p. 86 of Salem's book$^{1}$ about two particle correlations in the
ground state.

Electron correlations are manifested in the $n$-particle
reduced density matrices $\rhosN$ which, for the grand canonical
ensemble at inverse temperature $\beta$, are defined by
$$\rhosN (x_1,\ldots ,x_n; y_1,\ldots y_n)
= Z^{-1} \Tr \left[\cd{x_1 \sigma}\cd{x_2 \sigma}
  \ldots \cd{x_n \sigma}\c{y_n \sigma}\ldots
  \c{y_2 \sigma}\c{y_1 \sigma}\ebH\right].$$
The trace is over the full Fock space containing all particle numbers
$0$ to $2\abs\Lambda$,  and the normalization corresponds to
$\Tr \ \rhosN = \Tr [N_\sigma(N_\sigma-1)\cdots(N_\sigma-n+1)\ebH]/Z$,
where $N_\sigma$ counts the number of spin $\sigma$ particles.

Diagonalizing the hopping matrix through a change of basis, the Hamiltonian
becomes $H = \sum\nolimits_{\lambda} \lambda \bd \lambda \b \lambda.$
Here $\lambda$ runs over $2\abs\Lambda$ values.
$\bdl{}$ and $\bl{}$ are quasi-particle creation and annihilation
operators,  and are related to $\cd {x\sigma}$ and $\c{x\sigma}$ through
a unitary matrix $\phi$:
$$\cd {x\sigma} = \sum\limits_\lambda \phi^\lambda(x,\sigma)^* \bdl{} \qquad
  \c {x\sigma} = \sum\limits_\lambda \phi^\lambda(x,\sigma)   \bl{}.\eqno(9)$$
Consequently, $\bdl{}$ and $\bl{}$ obey the fermion anticommutation relations
$$\acomm{\bdl{}},{\b\mu }= \delta_{\lambda,\mu} \qquad
  \acomm{\bdl{}},{\bd\mu}=\acomm{\bl{}},{\b\mu}=0.$$
Of course,  $\rhosN$ can be expressed in terms of $\bdl{}$ and $\bl{}$ via (9),
and it is in this form that it will be most easily evaluated.

Letting $\nl{}=\bdl{}\bl{}$, the anticommutation relations imply
$$\Trace {\bdl1\cdots\bdl n\bm n\cdots\bm1} = \detij
    {\delta_{\lambda_i \mu_j}} \Trace {\nl 1\cdots\nl n}\eqno(10)$$
since the left hand side vanishes unless the $\lambda_i$ are distinct and
$\mu_i = \lambda_{\pi(i)}$ for some permutation $\pi \in S_n$ on
$n$ labels.  The antisymmetry is manifested in the determinant.
For the grand canonical ensemble,  the trace is over the full
Fock space,  which is an (antisymmetrized) tensor product of the one
particle Hilbert space.  Taking the trace in the $\bdl{}$ basis, $H$
is diagonal so
$$Z^{-1}\Trace {\nl 1\cdots\nl n} =
          \prod_{i=1}^n (1+e^{\beta\lambda_i})^{-1}.$$
Because this is a product,  it can be absorbed into the determinant in (10).
The right hand side of (10)
then becomes $Z\, \detij {\delta_{\lambda_i \mu_j} \Boltzf i}$.
As a particular instance,
$Z^{-1}\Trace {\bdl{} \bm {}} = \delta_{\lambda \mu} \Boltzf{}$, thus,
$$Z^{-1}\Trace {\bdl1\cdots\bdl n\bm n\cdots\bm1} =
    \detij {Z^{-1}\Trace {\bdl i \bm j}}.$$
Multiplying both sides by
$\phi^{\lambda_1}(x_1,\sigma)^* \cdots \phi^{\lambda_n}(x_n,\sigma)^*
 \phi^{\mu_1}(y_1,\sigma)         \cdots \phi^{\mu_n}(y_n,\sigma)$
and summing over all $\lambda$'s and $\mu$'s brings us back to the
position basis through (9).  Since the determinant is termwise multilinear
in the $\bdl i$ and $\bm j$,  we obtain Wick's Theorem:
$$
\rhosN (\{x_i;y_i\}_{i})
= {\rm det}\left[ \rhobes^{(1)}(x_i, y_j) \right]_{i,j=1}^n.$$

Now take all $x_1, \ldots, x_n, y_1, \ldots, y_n$ in the same sublattice,
either A or B.  By the Uniform Density Theorem,
$$\rhosN (\{x_i;y_i\}_{i})
= 2^{-n} \deltadet.$$
Explicitly,
$\rhosN (x_1,\ldots ,x_n; y_1,\ldots y_n) = \pm 2^{-n}$
when the $x_i$ are distinct and
$x_i=y_{\pi(i)}$ for some
fixed permutation $\pi \in S_n$. The sign of $\rhosN$ is
selected by the parity of $\pi$. Otherwise, $\rhosN = 0$.

This is to say that there can be no spatial correlations in the electron
density between sites of the same sublattice,  save only that the
probability of finding more than one electron in the same site and
spin state vanishes.  As an example, if $x,y \in A$ (or $B$) the
$\sigma$-electron pair density is
$\rhobes^{(2)} (x,y;x,y) =
(1-\delta_{xy})/4,$
the product of the one-particle densities at $x \not= y$.  Thus the
Pauli pressure is not felt between the electrons at these sites.
This is surprising since one generally expects,
and indeed will find, correlations between arbitrary
sites $x \in A$ and $y \in B$.

The extension to Hamiltonians of the form (8) is immediate.
The $n$-particle reduced density matrices now depend on spin as
well as spatial co-ordinates,  and are defined in the obvious way.
Taking all the $(x_i, \sigma_i)$ and $(y_i, \tau_i)$ to be in the
same sublattice,  the conclusion is that
$$\rhoN(x_1, \sigma_1,\ldots,x_n, \sigma_n; y_1, \tau_1, \ldots, y_n, \tau_n)
= 2^{-n} \deltadett.$$

Unfortunately,  these results extend neither to the canonical ensemble
nor to the interacting Hamiltonian.  For the canonical ensemble,
correlations result from the fixed number of available particles.
To see this,  consider a six site lattice in which four of the sites
are totally isolated,  and the remaining two are joined by a a non-zero
hopping matrix element~$t$.  The hopping matrix eigenvalues are $0$ with
multiplicity four,  and $\pm \abs t$.  If this lattice is populated with
spinless fermions, then each of the 16 ground states will involve one
particle hopping between the connected
sites,  and zero to four particles distributed over the isolated sites.
However, only 6 of these states,  those having a total of three particles,
will be represented in the canonical Gibbs state.
Select any two of the isolated sites,  and consider the probability of
finding them both occupied.  In the grand canonical ensemble,  we
have seen that this must be $1/4$ at any temperature.  However,  in the
zero temperature canonical ensemble,  the likelihood of this event is
${\displaystyle {4 \choose 2}}^{-1}=1/6$.  Of course,  this is a finite
size effect.  On the other hand, it is easy to create
transparent counterexamples for the interacting case by
taking $T=0$.  In particular,  the Ising model
Hamiltonian $H=\sum_{i,j \in L} J_{ij}s_is_j$  can be embedded in the
interaction term of (3).  The identifications are
$L=\Lambda \times \spinstates$, $s_i = 2n_{x\sigma}-1$ where
$i=(x,\sigma)$,  and $J_{ij} = U_{xy\sigma\tau}$
so that an occupied spin-site state corresponds to an up Ising spin.
This model is well known to have correlations between all sites for a
connected lattice.

Actually,  we have just discussed two trivial limits for the
standard Hubbard model in which $U_{xy\sigma\tau}=U \delta_{xy}$.
The $U \to 0$ limit is the free electron model,  while
$U \to \infty$ (or equivalently $T \to 0$) corresponds to a
(spatially) disconnected Ising model.  In both of these limits all
spatial correlations which are detectable on a sublattice must vanish ---
at least for finite lattices and $\beta<\infty$,  where the observables
are continuous functions of $U$.  In particular,  the antiferromagnetic
correlations (which can be computed to be present on a small lattice,
and are conjectured to be long range for $D \geq 3$ dimensional square
lattices) are clearly
seen to be a result of interference between the kinetic and potential
terms in the Hamiltonian.

\bigskip
{\bf II.  The Falicov-Kimball model$^{10}$:}  In this model there is one
species of mobile, spinless electrons and one species of arbitrarily
fixed particles.  It is the same as the Hubbard model with the choice
$t_{xy\downarrow} \equiv 0$ and $t_{xy \uparrow} = t_{xy}$.  The
Uniform Density Theorem applies to this model as a special case.
Note that it says that {\it both} the mobile and immobile particles
have density 1/2.

\bigskip
{\bf III.  Ground States:}  If we define the ground state
$\uprho_\sigma$ as the limit $\beta \rightarrow \infty$ of the
canonical $\widetilde{\uprho}_{\beta\sigma}$, then (5) and (6) apply
there, too.  (Note:  For the interacting Hamiltonian,  it is not clear
that $\rhobe=\widetilde{\uprho}_\beta$ in the $\beta \to \infty$ limit.
Certainly the canonical ensemble and the grand canonical ensemble do
not coincide in this limit,  even for the free electron model.  In
Remark I. above, we
gave a contrived example involving a six site lattice
for which $\rhobe^{(2)} \ne \widetilde{\uprho}^{\;(2)}_\beta$ at
$\beta \to \infty$, in the process of showing that certain
correlations did not vanish in the canonical Gibbs state.) Of course,
the ground state is not generally unique, and there are other
possibilities for $\uprho_\sigma$, in which case (5) and (6) apply to
states that are invariant under $Y = WJ$.  Not every ground state is
$Y$ invariant.  In the case of the usual Hubbard model on a connected
lattice with $U > 0$ and real $Y$ it is known$^{11}$ that the ground state
has spin angular momentum $S = (\vert A \vert - \vert B \vert)/2$ and
it is unique apart from the $(2S+1)$-fold degeneracy associated with
$S^z = \mfr1/2 (N_\uparrow - N_\downarrow) \in \{ -S, -S+1, \dots ,
+S\}$.  Ground states that are $Y$ invariant are mixtures of states
with $S^z$ and $-S^z$, i.e., $\mfr1/2 \vert S^z \rangle \langle S^z
\vert + \mfr1/2 \vert -\!S^z \rangle \langle -S^z \vert$ in Dirac's
notation.  The reason we can be sure that $Y \vert S^z \rangle = \vert
-\!S^z \rangle$ is this:  $Y \vert S^z \rangle$ is --- in any event ---
a state with $\mfr1/2 (N_\uparrow - N_\downarrow) = -S^z$.  It is also
a ground state.  By uniqueness, this state must be $\vert -\!S^z
\rangle$.  In the general model (3), we cannot be sure that spin flip
$=$ hole-particle transformation.

\bigskip
{\bf IV.  Infinite Volume Gibbs States:}  These states can, of course, have
different properties from finite volume states.  One way to define them is as
limits of finite volume states with specified boundary conditions that
need not respect $Y$ symmetry.  One example concerns the
Falicov-Kimball model with $U > 0$ on a hypercubic lattice in $D$
dimensions:  For $D \geq 2$ it has long-range order in the ground state
and at low temperatures, in which the up spins preferentially occupy
the $A$-sites and the down spins the $B$-sites (or vice versa)$^{12,13}$.
Similarly, the usual Hubbard model is expected to show the same
behavior when $D \geq 3$ if $U > 0$, at least if $U > 0$ is large
enough.  This has not been proved, however.  These examples violate
(5), but they do suggest that the charge density satisfies
$\uprho_{\beta \uparrow} (x,x) + \uprho_{\beta \downarrow} (x,x) = 1$
in these models.  If we now introduce nearest neighbor repulsion (which
is allowed in our general model) then even this constancy of the charge
density might be violated, however.

What is significant about our finite volume theorem (5), (6) --- {\it and
which does remain true in the infinite volume limit} --- is that for
every state with non-constant (spin or charge) density there is another
Gibbs state corresponding to the same temperature which has the
 complementary density.  In other words, {\it one
cannot invent a non-translationally invariant Hamiltonian} (either by
altering the hopping matrix or the pair potentials) with the property
that it {\it forces the density to increase in some specified regions
and to decrease in others} --- even though one might have thought {\it
a-priori} that the density can be controlled by the hopping or
potential energy.  Any attempt to cause a non-constant density will
always result in the certainty that {\it exactly} the reverse of the
desired non-constancy will occur with the hole-particle reversed
boundary condition.

Partial support by U.S. National Science Foundation grants
PHY90-19433A01 (EHL), DMS92-07703 (ML),  and a Canadian Natural
Sciences and Engineering Research Council 1967 Scholarship (RJM) is
gratefully acknowledged.  The authors wish to thank D.~Fisher,
D.J.~Klein and Z.G.~Soos for various fruitful discussions.

\bigskip\noindent
{\bf REFERENCES}
\item{$^{1}$} L. Salem, The Molecular Orbital Theory of Conjugated Systems
(New York: W. A. Benjamin, Inc.) 1966.
\item{$^{2}$} E. H\"uckel, Z. Phys. {\bf 70}, 204 (1931); {\bf 72}, 310
(1931); {\bf 76}, 628 (1932).
\item{$^{3}$}  C.A. Coulson and G.S. Rushbrooke, Proc. Camb. Phil. Soc.
{\bf 36}, 193 (1940).
\item{$^{4}$}  A.D. MacLachlan, Mol. Phys. {\bf 2}, 271 (1959).
\item{$^{5}$}  A.D. MacLachlan, Mol. Phys. {\bf 4}, 49 (1961).
\item{$^{6}$} R. Pariser and R.G. Parr, J. Chem. Phys. {\bf 21}, 466,
767 (1953).
\item{$^{7}$} J.A. Pople, Trans. Faraday Soc. {\bf 49},  1375 (1953).

\item{$^{8}$}  G.C. Wick, Phys. Rev. {\bf 80}, 268 (1950).
\item{$^{9}$}  M. Gaudin, Nuclear Phys. {\bf 15}, 89 (1960).
\item{$^{10}$}  L.M. Falicov and J.C. Kimball, Phys. Rev. Lett. {\bf 22}, 997
(1969).
\item{$^{11}$}  E.H. Lieb, Phys. Rev. Lett. {\bf 62}, 1201 (1989).  Errata {\bf
62}, 1927 (1989).  For an extension to positive temperature see K. Kubo and
T. Kishi, Phys. Rev. {\bf 62}, 1201 (1989).
\item{$^{12}$}  T. Kennedy and E.H. Lieb, Physica {\bf A138}, 320 (1986).
\item{$^{13}$}  U. Brandt and R. Schmidt, Z. Phys. {\bf B67}, 43 (1986).

\end